\documentstyle[emulateapj]{article}
\submitted{Submitted to Ap.J. Letters}
\def\gsim{\;\rlap{\lower 2.5pt
 \hbox{$\sim$}}\raise 1.5pt\hbox{$>$}\;}
\def\lsim{\;\rlap{\lower 2.5pt
   \hbox{$\sim$}}\raise 1.5pt\hbox{$<$}\;}

\def\spose#1{\hbox to 0pt{#1\hss}}
\newcommand\beq{\begin{equation}}
\newcommand\eeq{\end{equation}}

\makeatletter

\newenvironment{figurehere}
  {\def\@captype{figure}}
  {}
\makeatother

\lefthead{HAIMAN \& LOEB}
\righthead{X-RAYS FROM QUASARS}
\begin{document}\begin{flushright}
{\footnotesize
FERMILAB-Pub-99/116-A}
\end{flushright}
\nopagebreak
\vspace{-\baselineskip}
\title{X--Ray Emission from the First Quasars}
\author{Zoltan Haiman}
\affil{NASA/Fermilab Astrophysics Center\\ Fermi National Accelerator
Laboratory, Batavia, IL 60510, USA, email: zoltan@fnal.gov}
\and
\author{Abraham Loeb}
\affil{Harvard-Smithsonian Center for Astrophysics\\
60 Garden St., Cambridge, MA 02138, USA, email: aloeb@cfa.harvard.edu}

\begin{abstract}

It is currently unknown whether the Universe was reionized by quasars or stars
at $z\ga 5$.  We point out that quasars can be best distinguished from stellar
systems by their X-ray emission.  Based on a simple hierarchical CDM model, we
predict the number counts and X-ray fluxes of quasars at high redshifts.  The
model is consistent with available data on the luminosity function of
high-redshift quasars in the optical and soft X-ray bands.  The cumulative
contribution of faint, undetected quasars in our model is consistent with the
unresolved fraction of the X-ray background.  We find that the {\it Chandra
X-ray Observatory} might detect $\sim10^{2}$ quasars from redshifts $z\ga 5$
per its $17^\prime\times17^\prime$ field of view at the flux threshold of $\sim
2\times 10^{-16}~{\rm erg~s^{-1}~cm^{-2}}$. The redshifts of these faint
point-sources could be identified by follow-up infrared observations from the
ground or with the {\it Next Generation Space Telescope}.

\end{abstract}


\section{Introduction}

Cold dark matter (CDM) cosmologies predict that the first baryonic objects
appeared near the Jeans mass ($\sim 10^6~{\rm M_\odot}$) at redshifts as high
as $z\sim30$, and larger objects assembled later (Haiman \& Loeb 1999b, and
references therein).  It is natural to identify these objects as the sites
where the first stars or quasar black holes formed.  Observationally, bright
quasars are currently detected out to $z\sim 5$ (Fan et al. 1999), and galaxies
out to $z\sim 6.7$ (Chen et al. 1999; Weymann et al. 1998).  Although the
abundance of optically and radio bright quasars declines at $z\gsim 2.5$
(Schmidt et al. 1995; Pei 1995; Shaver et al. 1996), simple models based on the
Press--Schechter formalism reproduce this decline and simultaneously predict a
population of low-luminosity quasars that are too faint to be detected in
current surveys (Haiman \& Loeb 1998; Haehnelt et al. 1998).  Preliminary
evidence for such a population might already be indicated by the X--ray
luminosity function (LF) of quasars recently measured by ROSAT (Miyaji et
al. 1998a). The X--ray data probes fainter quasars than the optical data does,
and has not revealed a decline in the abundance of high-redshift quasars as
found in the optical (e.g. Schmidt et al. 1995).

The current census of high-redshift galaxies provides an estimate of the global
evolution of the star formation rate (SFR) in the redshift range $0\leq z\lsim
5$ (Madau 1999). Nevertheless, there is still considerable uncertainty about
the evolution of the SFR at the highest redshifts probed ($z\gsim 3$).  Recent
sub--mm observations from SCUBA indicate the existence of a substantial
population of dust--obscured star--forming galaxies that could raise the
implied SFR at high redshift (Barger et al. 1999, and references therein).
This uncertainty is particularly important, since it leaves open the question
of whether the observed population of stars produce sufficient ionizing
radiation by $z\sim5$ to reionize the intergalactic medium (IGM).  The
indication from the present data is that the SFR either declines or remains
roughly constant at $z\gsim 3$; however, in order to satisfy the Gunn--Peterson
constraint for the ionization of the IGM, the SFR needs to rise at high
redshifts above its value at $z\sim 3$ (Madau 1999).

The presence of a substantial population of faint quasars at $z\gsim 5$ is
therefore consistent with current data, and is suggested by cosmological models
for hierarchical structure formation.  If future data would indicate a
non-increasing cosmic SFR at $z\ga 3$, then quasars would be the likely sources
of reionization.  Recent theoretical models have focused on the properties of
the quasar population in the optical (Haehnelt \& Rees 1993; Haiman \& Loeb
1998; Haehnelt et al. 1998), and in the infrared (e.g. Sanders 1999; Almaini et
al.~1999). These models were shown to be only mildly constrained by the lack of
$z\gsim 3.5$ quasars in the Hubble Deep Field (HDF; see Haiman, Madau \& Loeb
1999).

In this {\it Letter} we point out that quasars can be best distinguished from
star forming galaxies at high redshifts by their X-ray emission.  First, we
illustrate the agreement between our simplest hierarchical model for quasars
and existing ROSAT data on the X--ray LF (Miyaji et al. 1998a) and the soft
X--ray background (XRB, Miyaji et al. 1998b).  We then show that forthcoming
X-ray observations with the {\it Chandra X-ray Observatory} ({\it CXO};
formerly known as {\it AXAF}) will be able to probe the abundance of quasars
during the reionization of the Universe at $z\ga 5$.

\section{Model Description and Quasar Spectra}

Our model is based on the Press--Schechter mass function of CDM halos (see
Haiman \& Loeb 1998 for complete details).  We assume that each halo of mass
$M_{\rm halo}$ forms a central black hole of mass $M_{\rm bh}$. The black hole
mass fraction $r\equiv M_{\rm bh}/M_{\rm halo}$ is assumed to obey a
log-Gaussian probability distribution
\beq 
p(r)=\exp [-(\log r -
\log r_0)^2/2\sigma^2]
\label{eq:scat}
\eeq 
with $\log r_0=-3.5$ and $\sigma=0.5$ (Haiman \& Loeb 1999b) These values
roughly reflect the distribution of black hole to bulge mass ratios found in a
sample of 36 local galaxies (Magorrian et al. 1998) for a baryonic mass
fraction of $\sim (\Omega_{\rm b}/\Omega_0)\approx 0.1$.  We further postulate
that each black hole emits a time--dependent bolometric luminosity in
proportion to its mass, $L_{\rm q}\equiv M_{\rm bh}f_{\rm q}=M_{\rm bh}L_{\rm
Edd} \exp(-t/t_0)$, where $L_{\rm Edd}=1.5\times10^{38}~M_{\rm bh}/{\rm
M_\odot}~{\rm erg~s^{-1}}$ is the Eddington luminosity, $t$ is the time elapsed
since the formation of the black hole, and $t_0=10^6$yr is the characteristic
quasar lifetime.  Finally, we assume that the shape of the emitted spectrum
follows the mean spectrum of the quasar sample in Elvis et al. (1994) up to a
photon energy of 10 keV.  We extrapolate the spectrum up to $\sim 50$ keV,
assuming a spectral slope of $\alpha$=0 (or a photon index of -1).

%*********************** FIGURE 1 **************************************
\vspace{+0.2cm}
\begin{figurehere}
%\epsscale{0.45} 
\plotone{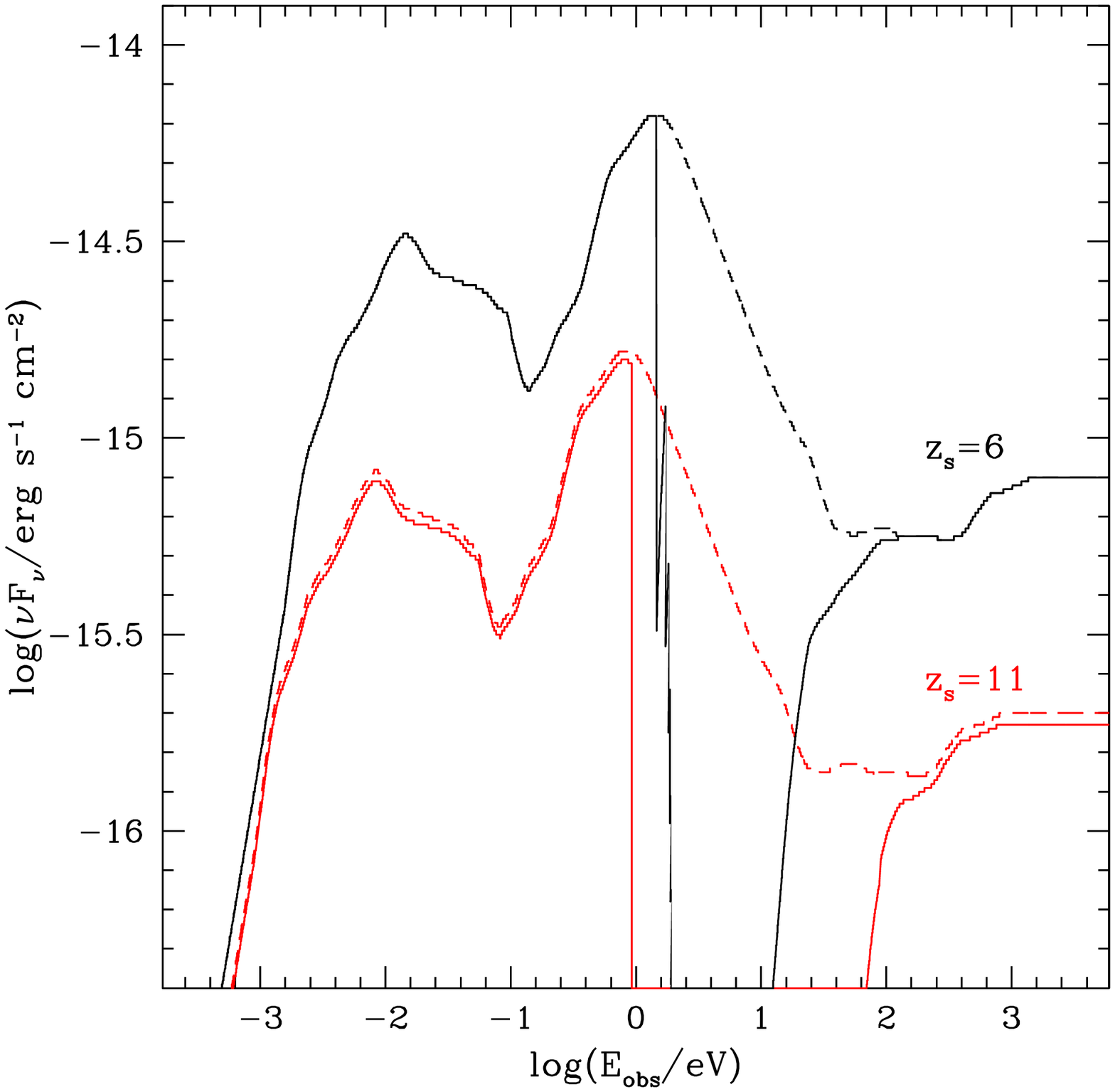}\vspace{-0.4cm}
\caption[Spectra] {\label{fig:spectrum} \footnotesize The observed spectra of
quasars with a central black hole mass of $M_{\rm bh}=10^8~{\rm M_\odot}$.  The
upper curves correspond to a source redshift of $z_{\rm s}=6$ and the lower
curves to a source redshift of $z_{\rm s}=11$.  In both cases we assume sudden
reionization at $z_{\rm r}=10$.  The dashed curves show the intrinsic spectral
shape, taken from Elvis et al. (1994).  We include the opacity of the neutral H
and He in the IGM at $z>z_{\rm r}$ and the opacity due to electron scattering
and the (extrapolated) Ly$\alpha$ forest in the IGM at $z<z_{\rm r}$.}
\end{figurehere}
\vspace{0.4cm}

This simple model was demonstrated to accurately reproduce the evolution of the
optical luminosity function in the B--band (Pei 1995) at redshifts $z\gsim 2.2$
(Haiman \& Loeb 1998).  Because our model incorporates several simplifying
assumptions, we regard it as the minimal toy model which successfully
reproduces the existing data. If one of our input assumptions was drastically
violated and our model had failed to fit the observed LF, then a modification
of its basic ingredients would be needed.  In this {\it Letter}, we focus on
the predictions of this minimal model in anticipation of the forthcoming launch
of {\it CXO}; an investigation of a broader range of plausible toy models will
be made elsewhere.  We adopt the concordance cosmology of Ostriker \&
Steinhardt (1995), namely a $\Lambda$CDM model with a tilted power spectrum
($\Omega_0,\Omega_\Lambda, \Omega_{\rm b},h,\sigma_{8h^{-1}},n$)=(0.35, 0.65,
0.04, 0.65, 0.87, 0.96).

Figure~\ref{fig:spectrum} shows the adopted spectrum of quasars, assuming a
black hole mass $M_{\rm bh}=10^8{\rm M_\odot}$, placed at two different
redshifts, $z_{\rm s}=11$ and $z_{\rm s}=6$.  In computing the intergalactic
absorption, we included the opacity of both hydrogen and helium as well as the
effect of electron scattering. We assumed that reionization occurred at $z_{\rm
r}=10$ and that at higher redshifts the IGM was homogeneous and fully
neutral. At lower redshifts, $0<z<z_{\rm r}$, we included the hydrogen opacity
of the Ly$\alpha$ forest given by Madau (1995), and extrapolated his fitting
formulae for the evolution of the number density of absorbers beyond $z=5$ when
necessary.  As Figure~\ref{fig:spectrum} shows, the minimum black hole mass
detectable by the $\sim 2\times 10^{-16}~{\rm erg~s^{-1}~cm^{-2}}$ flux limit
of {\it CXO} (see below) is $M_{\rm bh}\sim 10^8~{\rm M_\odot}$ at $z=10$ and
$M_{\rm bh}\sim 2\times10^7~{\rm M_\odot}$ at $z=5$.  In our model, the
corresponding halo masses are $M_{\rm halo}\sim 3\times10^{11}~{\rm M_\odot}$,
and $M_{\rm halo}\sim 6\times 10^{10}~{\rm M_\odot}$, respectively.  Although
such massive halos are rare, their abundance is detectable in wide-field
surveys.  Note that an accurate determination of the spectrum below
$\sim0.1$keV could have provided an estimate of the reionization redshift
$z_{\rm r}$.  Unfortunately, this spectral regime suffers from Galactic
absorption (O'Flaherty and Jakobsen 1997) and is outside the 0.4--6keV
detection band of {\it CXO}.

\section{The X--ray Luminosity Function}

In our model, the X--ray luminosity function at a redshift $z$ [in ${\rm
Mpc^{-3}~(erg/s)^{-1}}$] is given by a sum over halos that formed just before
that redshift,
\beq 
\phi(L_X,z) =\int_{0}^{t(z)} \frac{dt}{f_{\rm
X}f_{\rm q}(\Delta t)} \left.\frac{d^2N}{dM_{\rm bh}dt}\right|_{M_{\rm bh}=
\frac{L_X}{f_{\rm X}f_{\rm q}(\Delta t)}} \,\,\,
\label{eq:LF}
\eeq 
where $L_X$ is the observed X--ray luminosity in the instrument's detection
band ($0.5$--$3$ keV for ROSAT and 0.4--6 keV for {\it CXO}); $f_{\rm X}$ is
the fraction of the quasar's bolometric luminosity emitted in this band;
$d^2N/dM_{\rm bh}dt$ is the black halo formation rate, given by a convolution
of the Press--Schechter halo mass function with equation~(\ref{eq:scat}); and
$\Delta t=t(z)-t$ is the time elapsed from a cosmic time $t$ until a redshift
$z$.

Although our model was constructed so as to fit the observed optical LF,
Figure~\ref{fig:LF} demonstrates that it is also in good agreement with the
data on the X--ray LF. This implies that the choice of quasar spectrum in our
model is reasonable.  The solid curve in this figure shows the prediction of
equation~(\ref{eq:LF}) at $z=3.5$, near the highest redshift where X-ray data
is available.  The bottom curve corresponds to a cutoff in circular velocity
for the host halos of $v_{\rm circ}\geq 50~{\rm km~s^{-1}}$, which is
introduced here in order not to over--predict the number of quasars in the HDF
(Haiman, Madau \& Loeb 1999). The data points are from recent ROSAT
measurements, and the dashed curve in this figure shows a fitting formula from
Miyaji et al. (1998).  Note that the faintest quasar actually observed has
$L_X\lsim 10^{44}~{\rm erg~s^{-1}}$, and the fitting formula below this
luminosity is highly uncertain.  Nevertheless, it is important to remember that
our model was calibrated so as to fit the observed optical luminosity function
of quasars; the existence of a population of obscured quasars which are faint
in the optical band but bright in X--rays could increase the number counts
beyond our model predictions.

%*********************** FIGURE 2 **************************************
\vspace{+0.2cm}
\begin{figurehere}
%\epsscale{0.45} 
\plotone{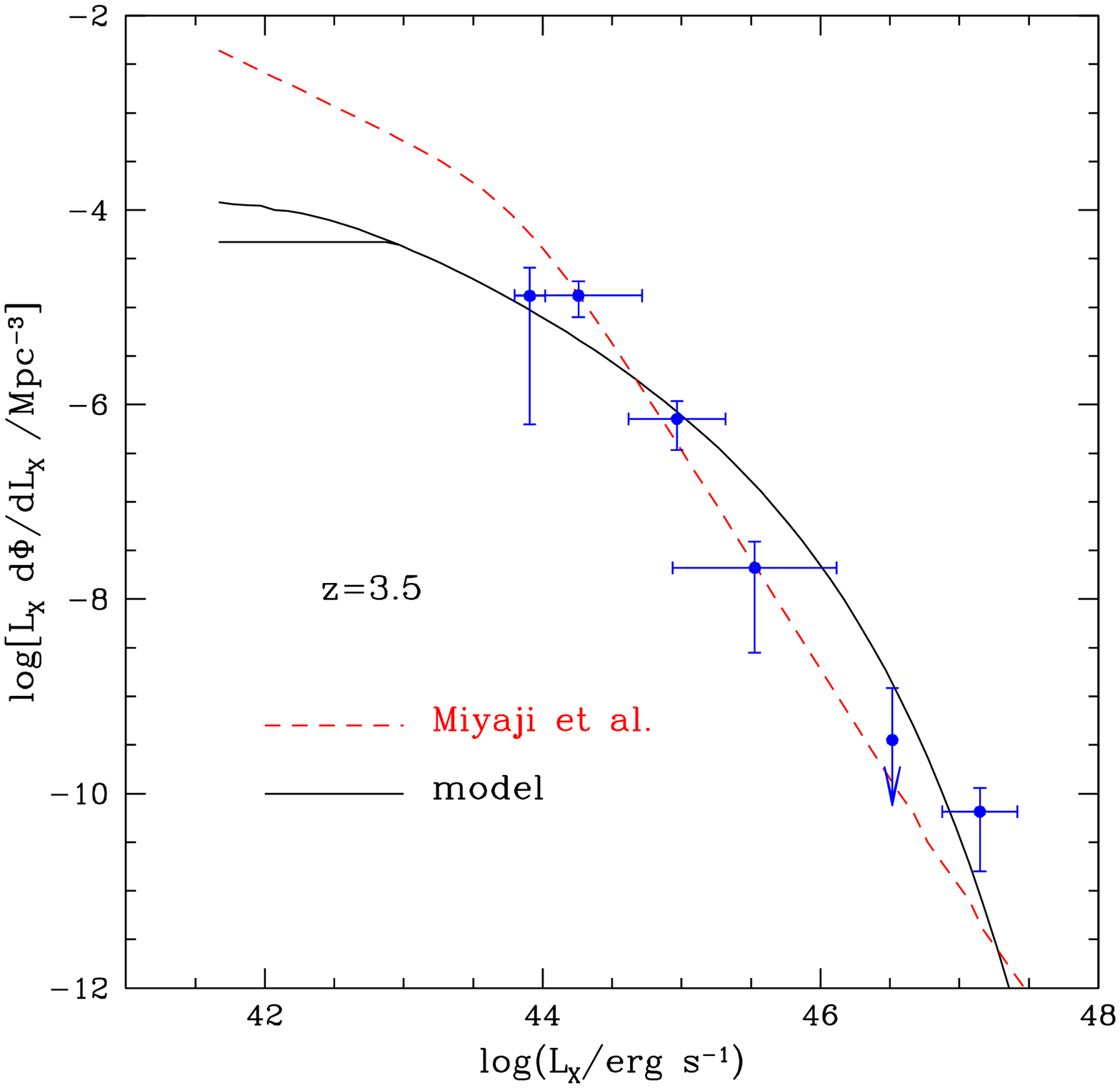}\vspace{-0.4cm}
\caption[Luminosity Function at z=3.5] {\label{fig:LF} \footnotesize The
predicted X--ray luminosity function at $z=3.5$ in our model (solid curves).
The lower curve shows the effect of a cutoff in circular velocity for the host
halos of $v_{\rm circ}\geq 50~{\rm km~s^{-1}}$. The ROSAT data points are
adopted from Miyaji et al. (1998a) and the dashed curve shows their fitting
formula (for our background cosmology).  }
\end{figurehere}
\vspace{0.1cm}

\section{The X--ray Background}

Existing estimates of the X--ray background (XRB) provide another useful check
on our quasar model.  The unresolved background flux at a photon energy $E$ is
given by (Peebles 1993)
\beq 
F(E)= c \int_0^\infty dz \frac{dt}{dz}
j(E_z,z)\,\,\,\,\,{\rm keV~cm^{-2}~s^{-1}~sr^{-1}~keV^{-1}},
\label{eq:xrb} 
\eeq 
where $E_z=E(1+z)$; and $j(E_z,z)$ is the comoving emissivity at a local photon
energy $E_z$, in units of ${\rm keV~cm^{-3}~s^{-1}~sr^{-1}~keV^{-1}}$, from
quasars shining at a redshift $z$.  This emissivity is a sum over all quasars
whose individual observed flux at $z=0$ is below the ROSAT PSPC detection limit
for discrete sources of $2\times10^{-15}~{\rm erg~cm^{-2}~s^{-1}}$ (Hasinger \&
Zamorani 1997).

Figure~\ref{fig:xrb} shows the predicted spectrum of the XRB in our model at
$z=0$ (solid lines).  In computing the background spectrum, we ignored the HI
absorption in the IGM, since it is negligible at energies above 100 eV.  We
also carried out the integral in equation~(\ref{eq:xrb}) only for $z>2$, the
redshift range where our model is valid (Haiman \& Loeb 1998).  The short
dashed lines show the predicted fluxes assuming a steeper spectral slope beyond
10 keV ($\alpha=-0.5$, or a photon index of -1.5). The long dashed line shows
the 25\% unresolved fraction of the soft XRB observed with ROSAT (Miyaji et
al. 1998b; Fabian \& Barcons 1992).  This fraction represents the observational
upper limit on the component of the soft XRB that could in principle arise from
high-redshift quasars.  As the figure shows, our quasar model predicts an
unresolved flux just below this limit in the 0.5-3 keV range.  The model also
predicts that most ($\gsim 90\%$) of this yet unresolved fraction arises from
quasars beyond $z=5$.

%*********************** FIGURE 3 **************************************
\vspace{+0.2cm}
\begin{figurehere}
%\epsscale{0.45} 
\plotone{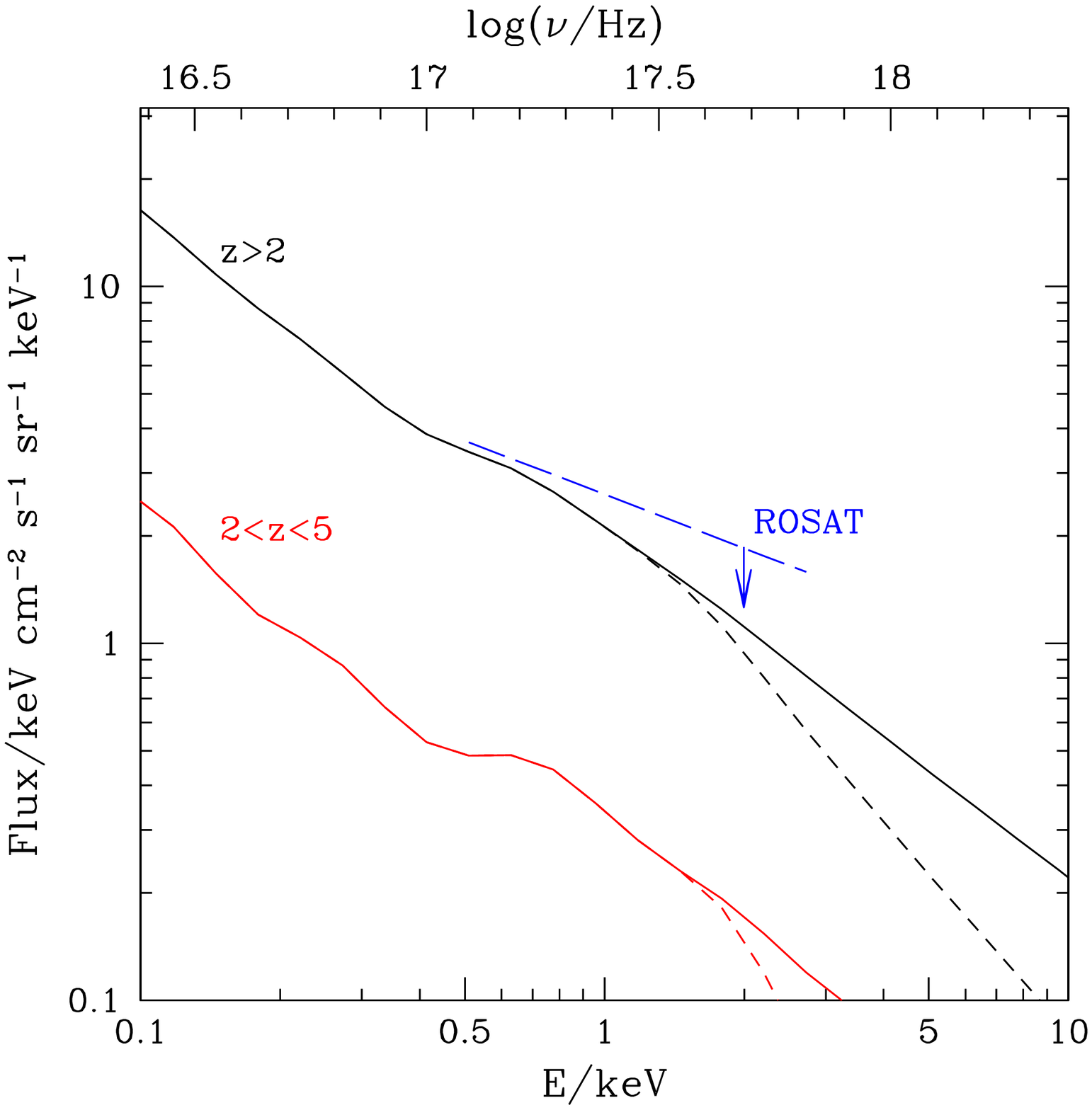}\vspace{-0.4cm}
\caption[X--Ray Background] {\label{fig:xrb} \footnotesize Spectrum of the soft
X--ray background in our model.  We assume that the median X-ray spectrum of
each source follows the mean spectrum of Elvis et al. (1994) up to 10 keV, and
has a spectral slope of 0.5 (solid lines) or -0.5 (short--dashed lines) at
higher photon energies.  The lower curves show the spectra resulting from
quasars with redshifts between $2<z<5$, and the upper curves include
contributions from all redshifts $z>2$.  In both cases, we include only quasars
whose individual fluxes are below the ROSAT PSPC limit of $2\times10^{-15}~{\rm
erg~cm^{-2}~s^{-1}}$ for the detection of a discrete source.  The long--dashed
line shows the unresolved fraction (assumed to be 25\%) of the soft X--ray
background spectrum from Miyaji et al. (1998b).}
\end{figurehere}
\vspace{0.1cm}

\section{Predicted Number Counts for the Chandra X-ray Observatory}

By summing the luminosity function over redshifts, we obtain the number counts
of quasars per solid angle expected to be detectable in a flux interval $dF_X$
around $F_X=L_X/4\pi d_L^2(z)$, for all sources above a redshift $z$:
\beq 
\frac{dN}{dF_X d\Omega} = \int_z^\infty dz
\left(\frac{dV}{dz d\Omega}\right) \phi(L_X,z) 4\pi d_L^2(z),
\label{eq:counts}
\eeq 
where $(dV/dzd\Omega)$ is the comoving volume element per unit redshift and
solid angle, and $d_L(z)$ is the luminosity distance at a redshift $z$.

In Figure~\ref{fig:counts}, we show the predicted counts from
equation~(\ref{eq:counts}) in the 0.4--6keV energy band of the CCD Imaging
Spectrometer (ACIS) of {\it CXO}.  Note that these curves are insensitive to
our extrapolation of the template spectrum beyond 10 keV. The figure is
normalized to the $17^\prime \times 17^\prime$ field of view of the imaging
chips.  The solid curves show that of order a hundred quasars with $z>5$ are
expected per field.  The abundance of quasars at higher redshifts declines
rapidly; however, a few objects per field are still detectable at $z\sim 8$.
The dashed lines show the results for a minimum circular velocity of the host
halos of $v_{\rm circ}\geq 100~{\rm km~s^{-1}}$, and imply that the model
predictions for the {\it CXO} satellite are not sensitive to such a change in
the host velocity cutoff.  This is because the halos shining at the {\it CXO}
detection threshold are relatively massive, $M_{\rm halo}\sim 10^{11}~{\rm
M_\odot}$, and possess a circular velocity above the cutoff. In principle, the
number of predicted sources would be lower if we had assumed a steeper spectral
slope.  However, the agreement between the LF predicted by our model at
$z\approx 3.5$ and that inferred from ROSAT observations would be upset by such
a change, and require a modification of the model that would in turn tend to
counter-balance the decrease in the predicted counts.

%*********************** FIGURE 4 **************************************
\vspace{+0.2cm}
\begin{figurehere}
%\epsscale{0.45} 
\plotone{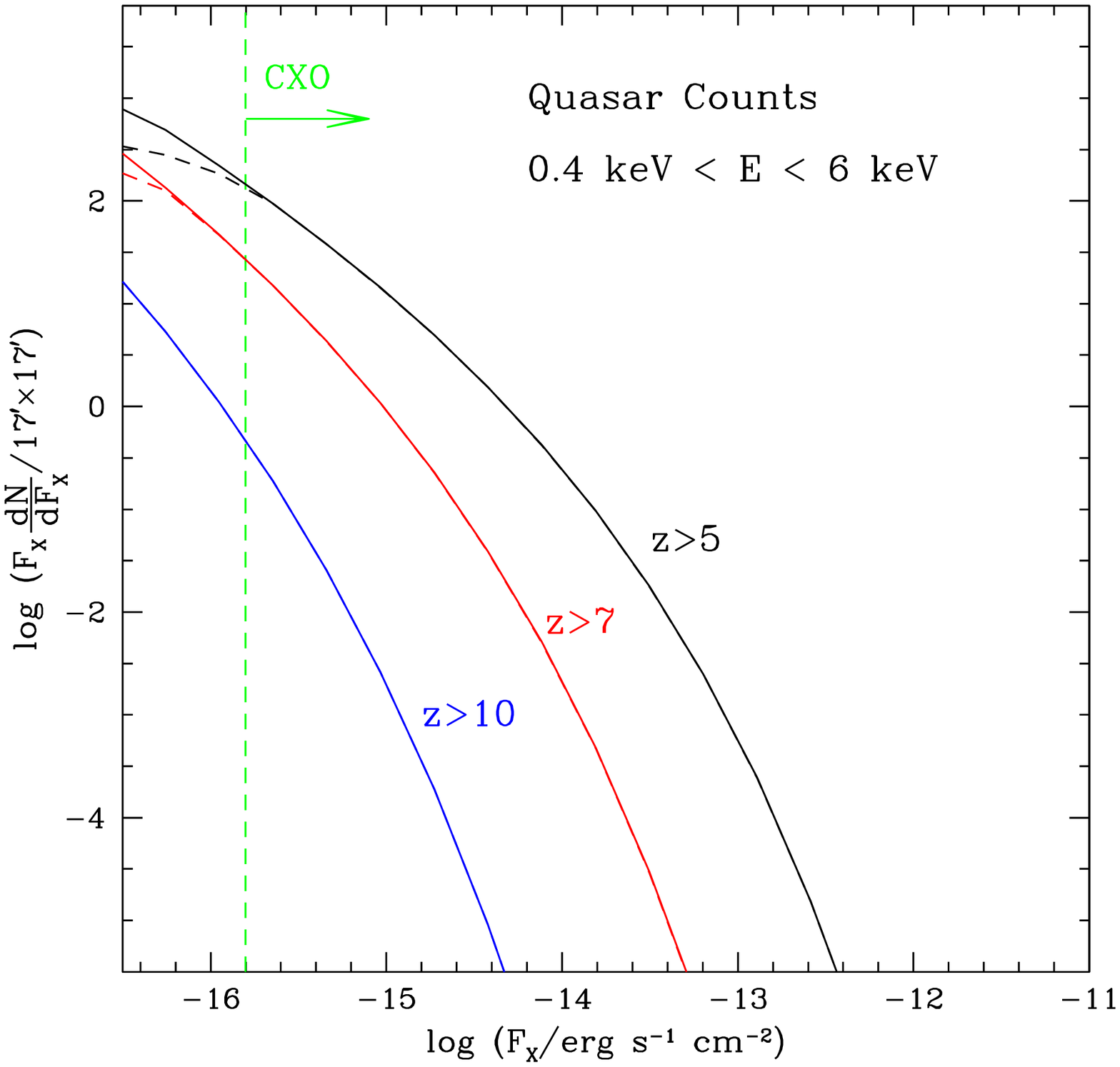}\vspace{-0.4cm}
\caption {\label{fig:counts} \footnotesize The total number of
quasars with redshift exceeding $z=5$, $z=7$, and $z=10$ are shown as a
function of observed X-ray flux in the {\it CXO} detection band.  The solid
curves correspond to a cutoff in circular velocity for the host halos of
$v_{\rm circ}\geq 50~{\rm km~s^{-1}}$, the dashed curves to a cutoff of $v_{\rm
circ}\geq 100~{\rm km~s^{-1}}$.  The vertical dashed line show the {\it CXO}
sensitivity for a 5$\sigma$ detection of a point source in an integration time
of $5\times10^5$ seconds.}
\end{figurehere}
\vspace{0.1cm}

\section{Discussion}

The existence of quasars at redshifts $z\ga 5$ has important consequences for
the reionization history of the Universe.  Quasars can be easily distinguished
from stellar systems by their X-ray emission.  We have demonstrated that
state--of--the--art X-ray observations could provide more stringent constraints
on quasar models than currently provided by the Hubble Deep Field (Haiman,
Madau, \& Loeb 1999).  In particular, we have found that forthcoming X--ray
observations with the {\it CXO} satellite might detect of order a hundred
quasars per field of view in the redshift interval $5\la z\la 8$.  Our
numerical estimates are based on the simplest toy model for quasar formation in
a hierarchical CDM cosmology, that satisfies all the current observational
constraints on the optical and X-ray luminosity functions of quasars.  Although
a more detailed analysis is needed in order to assess the modeling
uncertainties in our predictions, the importance of related observational
programs with {\it CXO} is evident already from the present analysis.  Other
future instruments, such as the HRC or the ACIS-S cameras on {\it CXO}, or the
EPIC camera on {\it XMM}, might also be useful in searching for high--redshift
quasars.

Follow-up optical and infrared observations are needed in order to identify the
redshifts of the faint point-like sources that might be detected by the {\it
CXO} satellite.  Quasars emit a broad spectrum which extends into the UV and
includes strong emission lines, such as Ly$\alpha$.  For quasars near the {\it
CXO} detection threshold, the fluxes at $\sim 1\mu$m are expected to be
relatively high, $\sim 0.5$--$0.8~\mu$Jy. Therefore, infrared spectroscopy of
X--ray selected quasars could prove to be a particularly useful approach for
unraveling the reionization history of the intergalactic medium at $z\ga 5$
(for the potential lessons to be learned, see Miralda-Escud\'e 1998; Haiman \&
Loeb 1999a; Loeb \& Rybicki 1999; and Rybicki \& Loeb 1999).

At present, the best constraints on hierarchical models of the formation and
evolution of quasars originate from the Hubble Deep Field.  However, {\it HST}
observations are only sensitive to a limiting magnitude of $V\sim29$ and cannot
probe the earliest quasars.  The Next Generation Space Telescope ({\it NGST}),
scheduled for launch in 2008, will achieve nJy sensitivity at wavelengths
$1-3~\mu{\rm m}$, and could directly probe the earliest quasars (Haiman \& Loeb
1998). The combination of X-ray data from the {\it CXO} satellite and infrared
spectroscopy from {\it NGST} could potentially resolve one of the most
important open questions about the thermal history of the Universe, namely
whether the intergalactic medium was reionized by stars or by accreting black
holes.

\acknowledgments

We thank G. Hasinger and R. Mushotzky for discussions that motivated this
study. We also thank N. White and M. Elvis for useful comments, and T. Miyaji
for supplying the data in Figure~\ref{fig:LF}.  ZH was supported by the DOE and
the NASA grant NAG 5-7092 at Fermilab. AL was supported in part by NASA grants
NAG 5-7039 and NAG 5-7768.

\end{document}